\documentclass[11pt,aps,tightenlines,nofootinbib,longbibliography,superscriptaddress]{revtex4-1}

\usepackage[utf8]{inputenc}
\usepackage{charter} % updated to match snowmass template
\usepackage{fullpage}
\usepackage{hyperref}
\usepackage{graphicx}
\usepackage{lipsum}
\usepackage{hyperref}
\hypersetup{hidelinks}
\usepackage[total={6.5in,9in},top=1in,headsep=0.1in,headheight=1in]{geometry}
\usepackage{enumitem,amssymb}
\usepackage[usenames,dvipsnames]{xcolor}
\usepackage{aas_macros}
\usepackage{xspace}
\usepackage{lineno}
\usepackage{siunitx}
\usepackage{booktabs}
\newcommand\snowmass{
\begin{center}
  \rule[-0.2in]{\hsize}{0.01in}\\
  \rule{\hsize}{0.01in}\\
  \vskip 0.1in
  Submitted to the Proceedings of the US Community Study\\ 
  on the Future of Particle Physics (Snowmass 2021)\\
  \rule{\hsize}{0.01in}\\
  \rule[+0.2in]{\hsize}{0.01in}\\[-2em]
\end{center}
}

\usepackage[firstpage=true]{background}
\backgroundsetup{contents={\parbox{6.5in}{\snowmass}}, scale=1,placement=top,opacity=1,color=black,position={3.25in,1.2in}}

\usepackage{fancyhdr}
\fancypagestyle{plain}{%
  \fancyhf{}%
  \fancyhead[C]{}
  \fancyfoot[C]{\thepage}
}

\fancypagestyle{empty}{%
  \fancyhf{}%
  \fancyhead[C]{{\it Snowmass2021: Future Gravitational-Wave Detector Facilities}}
  \fancyfoot[C]{\thepage}
}
\DeclareSIUnit{\ppm}{ppm}
\DeclareSIUnit{\torr}{torr}
\DeclareSIUnit{\clight}{\text{\ensuremath{c}}}
\DeclareSIUnit{\inch}{in}
\DeclareSIUnit{\atomicmassunit}{u}

\newcommand\aplusBnsHorizon{0.19}
\newcommand\aplusBbhHorizon{2.7}

\newcommand\aplusBnsSNR{75} % z = 0.01 
\newcommand\aplusEarlyWarning{4} % z = 0.01 
\newcommand\voyBnsHorizon{0.45}

\newcommand\ceTwoGlassBnsHorizon{8.3}
\newcommand\ceTwoGlassBbhHorizon{41}

\newcommand\ceTwoGlassBnsSNR{1.26e+03} % z = 0.01 
\newcommand\ceTwoGlassEarlyWarning{103} % z = 0.01 
\newcommand\ceTwoSiliconBnsHorizon{11.7}
\newcommand\ceTwoSiliconBbhHorizon{41}

\newcommand\ceTwoSiliconBnsSNR{1.46e+03} % z = 0.01 
\newcommand\ceTwoSiliconEarlyWarning{103} % z = 0.01 

\begin{document}

\title{Snowmass2021 Cosmic Frontier White Paper: Future Gravitational-Wave Detector Facilities}

\author{Stefan W. Ballmer}
\affiliation{Department of Physics, Syracuse University, Syracuse, NY 13244, USA}

\author{Rana Adhikari}
\affiliation{LIGO Laboratory, California Institute of Technology, Pasadena, CA 91125, USA}

\author{Leonardo Badurina}
\affiliation{Theoretical Particle Physics and Cosmology Group, Department of Physics, King’s College London, Strand, London, WC2R 2LS, UK}

\author{Duncan A. Brown}
\affiliation{Department of Physics, Syracuse University, Syracuse, NY 13244, USA}

\author{Swapan Chattopadhyay}
\affiliation{Fermi National Accelerator Laboratory, P. O. Box 500, Batavia, IL 60510, USA}

\author{Matthew Evans}
\affiliation{LIGO Laboratory, Massachusetts Institute of Technology, Cambridge, Massachusetts 02139, USA}

\author{Peter Fritschel}
\affiliation{LIGO Laboratory, Massachusetts Institute of Technology, Cambridge, Massachusetts 02139, USA}

\author{Evan Hall}
\affiliation{LIGO Laboratory, Massachusetts Institute of Technology, Cambridge, Massachusetts 02139, USA}

\author{Jason M. Hogan}
\affiliation{Department of Physics, Stanford University, Stanford, CA 94305, USA}

\author{Karan Jani}
\affiliation{Department Physics and Astronomy, Vanderbilt University, 2301 Vanderbilt Place, Nashville, TN, 37235, USA}

\author{Tim Kovachy}
\affiliation{Department of Physics, Stanford University, Stanford, CA 94305, USA}

\author{Kevin Kuns}
\affiliation{LIGO Laboratory, Massachusetts Institute of Technology, Cambridge, Massachusetts 02139, USA}

\author{Ariel Schwartzman}
\affiliation{SLAC National Accelerator Laboratory, Menlo Park, CA 94025, USA}

\author{Daniel Sigg}
\affiliation{LIGO Hanford Observatory, Richland, WA 99352, USA}

\author{Bram Slagmolen}
\affiliation{OzGrav, Research Schools of Physics, and of Astronomy and Astrophysics, Austrian National University, Canberra, ACT 2601, Australia}

\author{Salvatore Vitale}
\affiliation{Kavli Institute for Astrophysics and Space Research and Department of Physics, Massachusetts Institute of Technology, Cambridge, Massachusetts 02139, USA}
\affiliation{LIGO Laboratory, Massachusetts Institute of Technology, Cambridge, Massachusetts 02139, USA}

\author{Christopher Wipf}
\affiliation{LIGO Laboratory, California Institute of Technology, Pasadena, CA 91125, USA}

\begin{abstract}
The next generation of gravitational-wave observatories can explore a wide range of fundamental physics phenomena throughout the history of the universe. These phenomena include access to the universe’s binary black hole population throughout cosmic time, to the universe’s expansion history independent of the cosmic distance ladders, to stochastic gravitational-waves from early-universe phase transitions, to warped space-time in the strong-field and high-velocity limit, to the equation of state of nuclear matter at neutron star and post-merger densities, and to dark matter candidates through their interaction in extreme astrophysical environments or their interaction with the detector itself. We present the gravitational-wave detector concepts than can drive the future of gravitational-wave astrophysics. We summarize the status of the necessary technology, and the research needed to be able to build these observatories in the 2030s.
\end{abstract}

\maketitle

% Citations using natbib
%\citep{Snowmass2013:2013cqj,P5Report:2014pwa}

\newpage

\tableofcontents

\newpage

\section{Executive Summary}
\label{sec:exsum}
Gravitational-wave astronomy has revolutionized humanity's view of the universe. 
Investment in the field has rewarded the scientific community with the
first direct detection of a binary black hole merger~\cite{GW150914} and the multimessenger
observation of a neutron-star merger~\cite{LIGOScientific:2017vwq,MMA_BNS:2017}. Since the first detection of gravitational waves in 2015,
the National Science Foundation's LIGO and its partner
observatory, the European Virgo, have detected over ninety binary black hole mergers~\cite{LIGOScientific:2021djp}, a second neutron star merger~\cite{LIGOScientific:2020aai}, and evidence for neutron star--black hole binary mergers~\cite{LIGOScientific:2021qlt}. 

Major discoveries in astrophysics are driven by three related improvements:
better detector sensitivity, higher measurement precision, and opening new
observational windows. The next generation of gravitational-wave observatories promises all of these. 
%Gravitational-wave interferometers measure the gravitational-wave amplitude. Consequently, with a one order-of-magnitude sensitivity improvement
% over current detectors, 
The next-generation interferometers Cosmic Explorer~\cite{Reitze:2019iox,Evans:2021gyd} and Einstein Telescope~\cite{Punturo:2010zz} will see gravitational-wave sources across the history of the universe. The NEMO detector would compliment these facilities at higher frequencies~\cite{2019CQGra..36v5002H}. Atom interferometery~\cite{Abe_2021,canuel2018MIGA,Canuel2019ELGAR,ZAIGA2020,Badurina_2020} has the potential to open a new frequency window between the low-frequency Laser Interferometer Space Antenna (LISA) and ground-based laser interferometers. Next-generation gravitational-wave facilities will enable scientists to use the universe as a laboratory to test the laws of physics and study the nature of matter.  

As part of a multimessenger network of international gravitational-wave observatories,
astro-particle detectors, and telescopes across the electromagnetic spectrum~\cite{Snowmass-MMA,Snowmass-DMFac},
Cosmic Explorer and Einstein Telescope will precisely localize and study the
nature of a multitude of sources.  
Gravitational waves are generated by physical processes that are vastly
different from those that generate other forms of radiation and particles, and their
detections allow us to see into regions of the universe that cannot be observed in any other way. History tells us that it 
would be a profound anomaly in astronomy if nothing new and interesting came from
the vast improvement in sensitivity that next-generation gravitational-wave observatories will provide.

\begin{figure}[t]
\begin{center}
\includegraphics[width=1\textwidth]{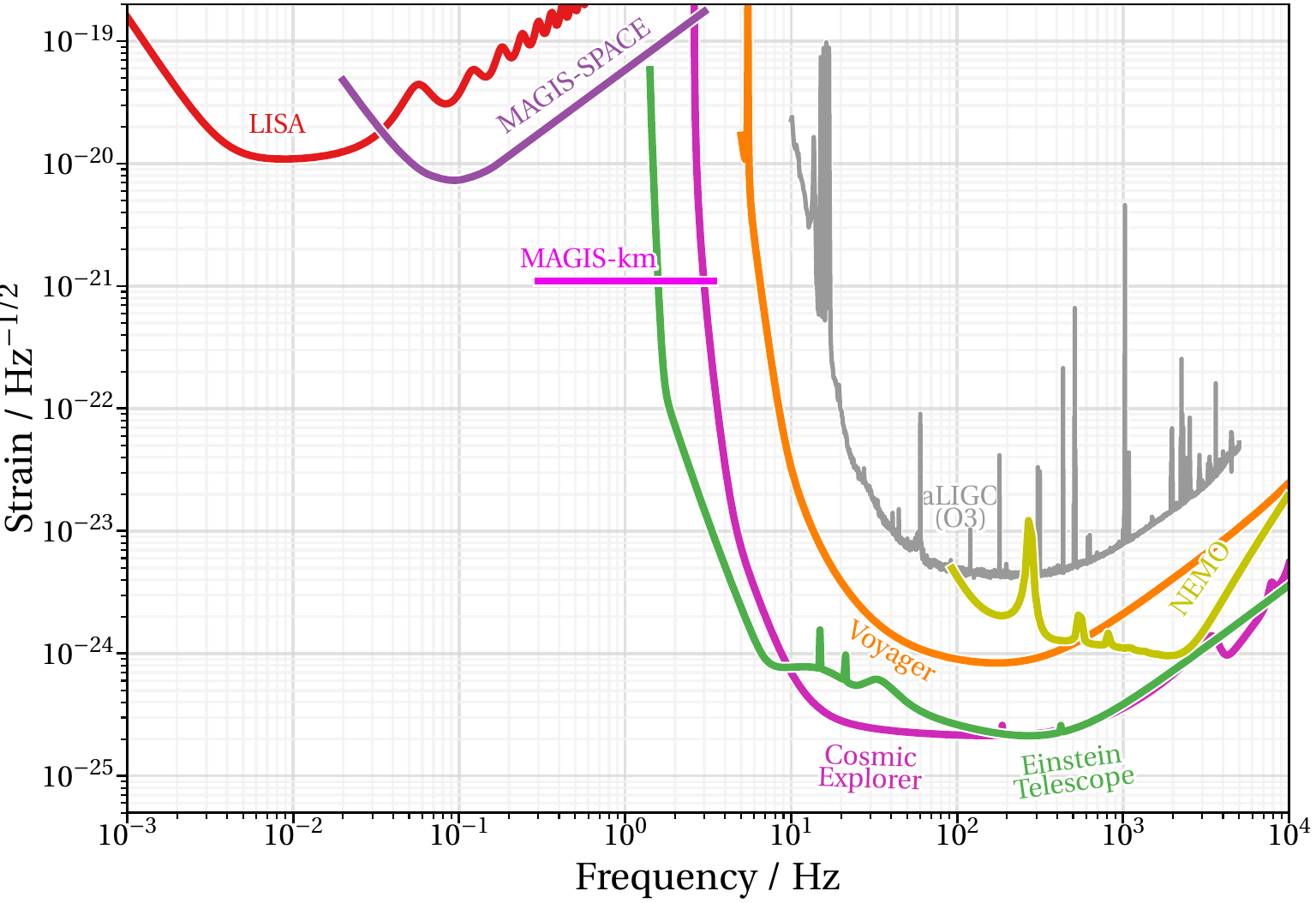}
\end{center}
    \caption{Amplitude spectral densities of detector noise for the next-generation laser interferometers Cosmic Explorer, LIGO Voyager, the proposed Australian NEMO detector, and the three paired detectors of the triangular Einstein Telescope. Detector noise curves are also shown for the proposed MAGIS-km atom interferometer and envisioned space-based follow-on detector (MAGIS-SPACE). The sensitivity curves of Advanced LIGO's last observation run (aLIGO O3) and of the Laser Interferometer Space Antenna (LISA) are shown for comparison.}
\label{fig:sense}
\end{figure}

Cosmic Explorer's increased sensitivity comes primarily from scaling up LIGO technology from 4~km to 40~km L-shaped arms, with the initial detector design focused on high sensitivity with low risk. The Einstein Telescope will use advanced detector technologies in a 10~km triangular interferometer with $60^\circ$ angles built underground to minimize low-frequency noise; multiple interferometers focus on different frequency domains and gravitational-wave polarizations.. A proposal known as LIGO Voyager would upgrade the existing LIGO facilities to the limit of their observational reach using advanced detector technologies~\cite{LIGO:2020xsf}. Although LIGO-Voyager does not reach the sensitivity of Cosmic Explorer and Einstein Telescope, it can be built in the existing LIGO facilities.  A comparison of the strain sensitivity of these proposed detectors, together with the LISA detector~\cite{Danzmann_1996}, the MAGIS-km\cite{Abe_2021} atom interferometer based on technology currently under development at Fermilab, and future space-based atom interferometers, is shown in Figure~\ref{fig:sense}.

%Finally, GLOC-Optimal is the projected sensitivity curve of a futuristic large moon-based laser interferometer.

This white paper complements other white papers that discuss the description of multi-messenger facilities~\cite{Snowmass-MMA} and observational facilities to study dark matter~\cite{Snowmass-DMFac}. Section \ref{sec:sci} reviews the scientific potential of future gravitational-wave facilities; we refer to the complementary white papers for detailed discussion~\cite{Snowmass-PBH,Snowmass-UDM,Snowmass-DMX,Snowmass-COS,Snowmass-FUN,Snowmass-DE,Snowmass-WF}. Section \ref{sec:gw} discusses the proposed U.S. laser interferometer detectors: Cosmic Explorer, a ground-based next-generation gravitational-wave observatory, and Voyager, a technology upgrade proposal to existing the gravitational-wave LIGO facilities. Section \ref{sec:ai} reviews atom interferometer technology, the status of the atom interferometers facilities MAGIS and AION, and opportunities for potential use of this technology in gravitational-wave detectors. We highlight a number of common technologies required for both laser and atom interferometers in section \ref{sec:ct}. Finally, we place terrestrial-based gravitational-wave detectors into the future landscape of observatories that can span the whole gravitational-wave spectrum (CMB B-modes, NANOgrav, LISA), and explore more speculative future possibilities, such as a potential lunar-based observatory.

\section{Introduction}
\label{sec:sci}

In the U.S., the proposed Cosmic Explorer observatory is designed to have ten times the sensitivity of Advanced LIGO and will push the reach of gravitational-wave astronomy towards the edge of the observable universe ($z \sim 100$)~\cite{Reitze:2019iox,Evans:2021gyd}. LIGO-Voyager can observe binary black holes with 30\,$M_\odot$ components to $z\approx7$. Understanding how the universe 
made the first black holes, and how these first
black holes grew, is one of the most important
unsolved problems in astrophysics.
Cosmic Explorer will see evidence for the first
stars by detecting the mergers of
the black holes they leave behind. The millions of binary mergers detected by Cosmic
Explorer will map the population of compact objects across time, detect
mergers of the
first black holes that contributed to seeding the universe's structure, 
explore the physics of massive stars, and reveal the processes that create black holes and neutron stars.

Binary neutron star mergers at cosmological distances will also be observable with Cosmic Explorer and LIGO Voyager. The maximum observable redshift for the A$+$ LIGO upgrade~\cite{DoublingRange:2015} is \aplusBnsHorizon, increasing to \voyBnsHorizon~in LIGO-Voyager. A network consisting of Cosmic Explorer in the U.S. and Einstein Telescope in Europe would detect $\gtrsim 10^5$ binary neutron star mergers per year, with a median redshift of $\sim$1.5---close to the peak of star formation---and a horizon of $z\gtrsim 9$~\cite{Borhanian:2022czq}. Approximately 200 of these binary neutron stars would be localized every year to better than one square degree, enabling followup with telescopes with small fields of view~\cite{Borhanian:2022czq}. The improved low-frequency sensitivity of third-generation detectors allows them to detect and localize sources prior to merger. The triangular configuration of the Einstein Telescope allows it to localize sources without a second observatory~\cite{Maggiore:2019uih}. The Einstein Telescope is able to detect 6 sources per year with more than 5 minutes before merger with a localization better than ten square degrees, some of which have up to 30 minutes warning time. A full three-detector network consisting of two Cosmic Explorer detectors and the Einstein Telescope would allow of order 10 sources per year to be localized localized to better than one square degree five minutes before the merger~\cite{Nitz:2021pbr}.

{\bf Neutron star equation of state:} By observing many hundreds of loud neutron star mergers and measuring the stars' radii
to $100$~m or better, next-generation detectors will probe the phase structure of
quantum chromodynamics, revealing the nuclear equation of state and its
phase transitions~\cite{2008PhRvD..77b1502F,2010PhRvD..81l3016H,2012PhRvD..85l3007D,2020GReGr..52..109C}. 
The ability of a gravitational-wave detector to study the hot,
dense remnants of neutron star mergers will provide an entirely new way
of mapping out the dense, finite-temperature region of the
quantum chromodynamics phase
space, a region that is currently unexplored~\cite{2009PhRvD..80l3009Y,2016PhRvL.117d2501K,2016EPJA...52...50O,2016PhRvL.117d2501K,2016EPJA...52...50O,2019PhRvL.122f1101M}.
Cosmic Explorer and LIGO Voyager are well suited to study the post-merger remnant~\cite{2018PhRvD..98d4044M,2018PhRvD..97b4049Y,2019PhRvD..99j2004M}. The proposed Australian  NEMO high-frequency third-generation detector~\cite{2019CQGra..36v5002H} would compliment the gravitatiational-wave picture of hot, dense matter. Together with multimessenger observations of the merger remnants, gravitational-wave observations will shape theoretical models describing fundamental many-body nuclear interactions and answer questions about the composition of matter at its most extreme, such as whether quark matter is realized at high densities~\cite{2019PhRvL.122f1102B,2019PhRvL.122f1101M,2020PhRvD.102l3023B}.
Next-generation detectors could distinguish between binaries containing boson stars from binary neutron stars and binary black holes by measuring the compactness parameter of the stars, which for boson stars typically lies between that of black holes and neutron stars~\cite{Sennett:2017etc,Toubiana:2020lzd}.

{\bf Cosmology:} Gravitational-wave standard sirens are expected to play an important role in the context of cosmology. Gravitational waves allow measurement of the luminosity distance of the source and, together with redshift measurements, can probe the distance-redshift relation~\cite{Schutz:1986gp}. Measurement of the Hubble parameter using standard sirens does not require a cosmic distance ladder and is model-independent: the absolute luminosity distance is directly calibrated by the theory of general relativity. Approximately fifty additional multi-messenger binary neutron star observations would be needed to resolve the tension between the Planck and R19 measurements of $H_0$ with a precision of 1--2$\%$~\cite{Nissanke:2013fka,Mortlock:2018azx}. The precision of third-generation detectors, combined with deep optical-to-near-infrared observatories, would allow third-generation observatories to resolve this tension\cite{Chen_2021}.

{\bf Fundamental tests of gravitation:} Next-generation facilities will enable high precision probes of highly curved spacetime~\cite{Will:2001, Yunes:2011}. LIGO's first observations of gravitational waves from binary black holes have already made it possible to perform the first tests of general relativity in the highly relativistic strong-field regime~\cite{GW150914:GR}. LIGO-Voyager and Cosmic Explorer will allow us to significantly improve the precision of such tests, as binary black hole mergers will be regularly detected with signal-to-noise ratios of hundreds to thousands. 
Observations of black holes could reveal violations of general relativity in the form of failure of the no-hair theorem as a result of quantum effects near black hole horizons~\cite{2014arXiv1409.7977S, 2015dvali}.  The physics that resolves the black hole information paradox could also give rise to postmerger gravitational wave echoes~\cite{2018PhRvD..98d4021C}. These echoes have not yet been observed~\cite{Cardoso:2016oxy, Mark:2017dnq, 2018PhRvD..98d4021C, Wang:2020ayy}, but Cosmic Explorer's extremely high sensitivity could reveal them, should they exist.

The vast cosmological distances---redshifts in excess of $z\sim 20$---over which gravitational waves travel, will severely constrain violation of local Lorentz invariance and the graviton mass~\cite{Will:2014kxa}. Such violations or a non-zero graviton mass would cause dispersion in the observed waves and hence help to discover new physics predicted by certain quantum gravity theories. At the same time, propagation effects could also reveal the presence of large extra-spatial dimensions that lead to different values for the luminosity distance to a source, as inferred by gravitational-wave and electromagnetic observations~\cite{Belgacem:2018lbp, Pardo:2018ipy}, or cause birefringence of the waves predicted in certain formulations of string theory~\cite{Alexander:2009tp, Alexander:2017jmt}. The presence of additional polarizations predicted in certain modified theories of gravity, instead of the two degrees of freedom in general relativity, could also be explored by future detector networks~\cite{Will:2014kxa, Isi:2017fbj}. 

{\bf Multi-messenger observations:} Multi-messenger observations of binary neutron star mergers are a promising new environment to probe weakly interacting light particles. Immediately after the merger, these remnants reach temperatures in the 30--100~MeV range and densities above $10^{14}~\text{g/cm}^3$, similar to the proto-neutron stars formed in core-collapse supernovae that have been used to place constraints on a wide range of scenarios. The large temperature and density of a post-merger remnant makes them very efficient at producing feebly interacting dark sector particles, which can escape this environment and lead to observational signals~\cite{Dietrich:2019shr,Harris:2020qim,Diamond:2021ekg}. Dark photons with masses in the 1--100~MeV range would be copiously produced and, for a large range of unconstrained couplings, would lead to a very bright transient gamma-ray signal originating from the dark photon decay~\cite{Diamond:2021ekg}. The precision and early warning offered by next-generation detectors allows the use of the associated gravitational-wave signal as a trigger and a timing measurement to help distinguish signal from background fluctuations and allows for gamma-ray observatories with narrower fields of view to observe events. Observations of gravitational waves from neutron star mergers can allow exploration of an object with a non-negligible contribution from vacuum energy to their total mass. The presence of vacuum energy in the inner cores of neutron stars occurs in new QCD phases at large densities, with the vacuum energy appearing in the equation of state for a new phase. This, in turn, leads to a change in the internal structure of neutron stars and influences their tidal deformabilities, which are measurable in the gravitational-wave signals of merging neutron stars~\cite{Csaki:2018fls}.

{\bf Dark Matter:} Light axions inside neutron stars can modify the binary inspiral yielding a dark matter signature detectable in the gravitational waveform. Constraints can be placed on axions with masses below $10^{-11}$eV and decay constants ranging from $10^{16}$~GeV to $10^{18}$~GeV, with next-generation facilities improving current constraints by a factor of $\sim 3$~\cite{Zhang:2021mks}. Spinning black holes can superradiantly amplify excitations in a surrounding field, generation long-lived ``bosonic clouds'' that slowly dissipate energy through the emission of gravitational waves~\cite{Arvanitaki:2014wva,Arvanitaki:2016qwi}. A large number of unresolved sources can contribute to the stochastic gravitational wave background, allowing Cosmic Explorer to constrain bosons in the mass range $\sim [7 \times 10^{-14}, 2 \times 10^{-11}]$~eV~\cite{Yuan:2021ebu}. In addition to direct detection, the boson cloud spins down the black hole to a characteristic spin determined by the boson mass and the black hole mass. With the large number of high signal-to-noise ratio events that will be seen by Cosmic Explorer, the existence of noninteracting bosons in the mass range $10^{-13}$ to $10^{-12}$~eV could be confirmed through their imprint on the black hole spin distribution~\cite{Ng:2019jsx}.

Scalar and vector ultralight dark can be probed through its direct interaction with gravitational-wave interferometers. Ultralight dark matter signatures can be detected in a laser interferometer through the changes in the optical-path-length difference between the two arms of the interferometer arise due to oscillations in the thickness of the beam-splitter and through oscillations in the refractive index of the beam-splitter material~\cite{Grote:2019DM-LIFO}. Searches for dark photons have been performed using data from the Advanced LIGO--Virgo observing runs~\cite{Pierce:2018xmy,Guo:2019ker,LIGOScientific:2021odm} and the GEO600 interferometer has been used to search for scalar, dilaton dark matter interactions~\cite{Dooley:2015fpa}. Next-generation gravitational-wave interferometers will enable deeper explorations of this space and ultralight dark matter that affects fundamental constants (such as the electron mass or the fine structure constant) will change the internal energy levels of the atoms in a way that can be explored by atom interferometers~\cite{Kuhn_2014,PhysRevLett.106.080801}.

{\bf Primordial cosmology:} Primordial black holes (PBHs) have been a longstanding candidate for some fraction of dark matter. Binaries containing containing sub-solar mass black holes or black hole binaries observed at very large red shifts would indicate the presence of PBHs as a dark matter component~\cite{Nakamura_1997,Bird:2016dcv} and next-generation facilities can constrain the fraction of dark matter in PBHs\cite{Chen_2020}. Primordial black holes could be surrounded by spikes of particle dark matter which could influence the dynamics of a binary merger allowing Cosmic Explorer to probe dark matter candidates with masses heavier than approximately $m_a \sim 10^{-6}$~eV~\cite{Bertone:2019irm}.
Gravitational waves can also be produces through a number of other mechanisms in the early universe, including inflation, phase transitions, topological defects, producing a
detectable signal and revealing information about fundamental physics~\cite{https://doi.org/10.48550/arxiv.2203.07972}.

{\bf Black hole formation history:} The evidence for the existence of intermediate-mass black holes
(IMBHs) in the $10^2$\,--\,$10^4\,M_\odot$ mass range is still inconclusive
at present.  Attempts to look for electromagnetic signatures are hampered by the small dynamical footprint of low-mass IMBHs and
the difficulty of associating phenomena such as ultraluminous x-ray sources specifically with IMBHs\cite{MillerColbert:2004}.  On the other
hand, a handful of promising sources have been observed\,\cite{Farrell:2009}, and multiple formation scenarios have been
proposed---though none without problems~\cite{Collaboration:S5HighMass}.  Gravitational-wave
observations of compact objects in this mass range, which would be
enabled by a future detector with good low-frequency sensitivity,
could yield the first definitive proof of IMBH existence at the low
end of the IMBH mass range\,\cite{Veitch:2015}.  Such measurements could also answer
outstanding questions about the dynamics of globular clusters and
about the formation history of today's massive black holes~\cite{Gair:2009ETrev}.

{\bf Supernovae:} Massive stars undergoing core-collapse supernova also generate gravitational waves from the dynamics of hot, high-density matter in their central regions. Cosmic Explorer and Einstein Telescope will be sensitive to supernovae within the Milky Way and its satellites, which are expected to occur once every few decades~\cite{2019PhRvD.100d3026S}. Core collapses should be common enough to have a reasonable chance of occurring during the few-decades-long lifetime of Cosmic Explorer. A core-collapse supernova seen by Cosmic Explorer will have a significantly larger signal-to-noise ratio than one seen by current gravitational-wave detectors, and could be detected by a contemporaneous neutrino detector like DUNE~\cite{2021EPJC...81..423A}, giving a spectacular multimessenger event. Detection of a core-collapse event in gravitational waves would provide a unique channel for observing the explosion's central engine~\cite{2020arXiv201004356A} and the equation of state of the newly formed protoneutron star~\cite{2018ApJ...861...10M}. Detection of a supernova would be spectacular, allowing measurement of the progenitor core's rotational energy and frequency measurements for oscillations driven by fallback onto the protoneutron star~\cite{2021PhRvD.103b3005A}.

Gravitational-wave memory may be left behind by most stellar collapse events, even
those that do not result in an explosion~\cite{Richardson:2021lib}. The typical growth timescale
of the memory is of order $\gtrsim 0.1\,\mathrm{s}$, which makes it
the only known low-frequency gravitational-wave emission process in stellar
collapse. Detecting the gravitational-wave memory from a galactic
event with Advanced LIGO may be a difficult task even if the full projected
low-frequency sensitivity is reached, but LIGO-Voyager and Cosmic Explorer would allow detection.

\section{New Laser Interferometer Gravitational-Wave Facilities and Upgrades}
\label{sec:gw}

%\subsection{Designing the Next-Generation Gravitational-Wave Observatories}
The LIGO and Virgo instruments have opened a new window on the universe. As discovery machines they were designed without knowledge of the actual signal population.
At its current sensitivity the Advanced LIGO detectors see a signal roughly once per week. When the ongoing ``A+'' upgrade is mature in 2025, they will deliver roughly ten detections per week.

The science questions presented in the preceding section are only addressable by making observations with significantly higher fidelity over a wider frequency band, and by observing more distant sources, driving the need for an order of magnitude greater sensitivity in the audio frequency band. This is achievable by either a significantly longer measurement baseline (detector arm length), a significant reduction in the dominant fundamental detector noise sources (most prominently quantum noise, thermal noise and seismic noise coupling), or ideally both. 

Cosmic Explorer is a third-generation detector design proposed as a successor to the second-generation Advanced LIGO detectors. It consists of two observatories, with at least one having a detector arm length of \SI{40}{\km}. Cosmic Explorer is described in detail in the horizon study published in 2021 \cite{Evans:2021gyd}. The design is summarized in section \ref{sec:ce}. Together with the European Einstein Telescope, Cosmic Explorer will ultimately span the globe with a three-detector next-generation network of gravitational-wave detectors.

LIGO Voyager is a cryogenic silicon gravitational-wave interferometer conceived as a possible upgrade to the LIGO $4~km$ facilities. It aims to significantly reduce the noise sources in Advanced LIGO by switching to cryogenic (123K), heavy silicon test masses, permitting lowering the thermal noise and increasing the laser power to reduce the quantum noise. Voyager is described in section \ref{sec:voy}.

\subsection{The Cosmic Explorer Observatory}
% [Stefan Ballmer]
\label{sec:ce}
The Cosmic Explorer observatory design was conceived as an extrapolation of the technology currently employed in Advanced LIGO. The entire facility is redesigned in order to optimize the observatory performance, enabling the science described in section \ref{sec:sci}. This section provides a technical overview of the Cosmic Explorer observatory. It also outlines the key technologies that will require research and development to enable the CE science goals. Finally, the key drivers of project costs are discussed.

\begin{figure}[!ht]
  \centering
  \includegraphics[height=0.6\textheight]{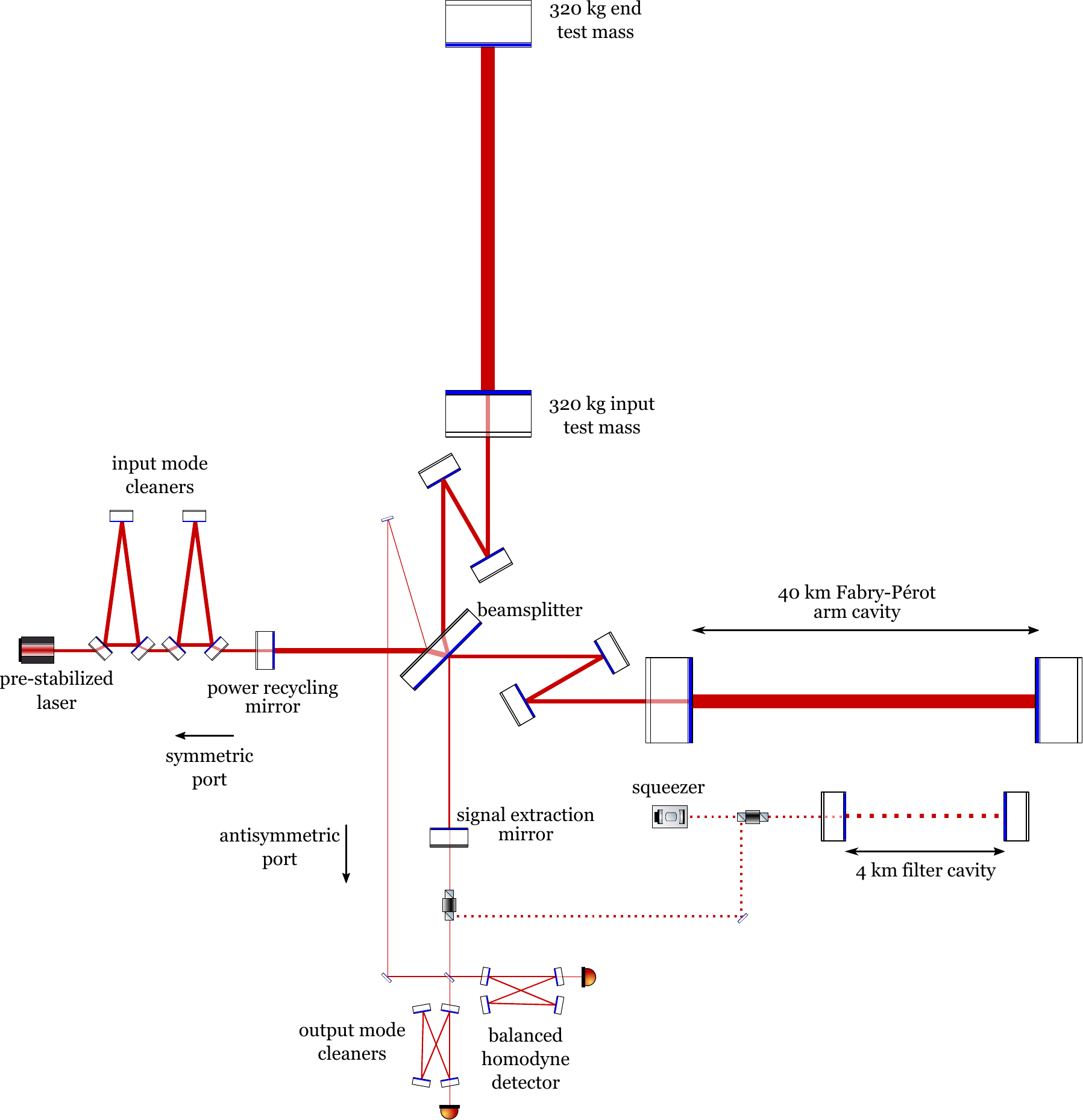}
  % \includegraphics[width=0.48\textwidth]{ce_design_simple.pdf}
  % \caption{Two options for a reference design figure.}
  \caption{Simplified optical layout of the Cosmic Explorer reference detector concept for the \SI{40}{\km} implementation. The input and end test masses form the two arm cavities which, together with the beamsplitter, power recycling mirror, and signal extraction mirror, comprise the core of the dual-recycled Fabry--P\'{e}rot Michelson interferometer. The light carrying the gravitational wave signal is spatially filtered and read out from the antisymmetric port by a balanced homodyne detector comprised of two photodiodes and output mode cleaners; a high power laser is injected into the symmetric port of the interferometer after passing through two input mode cleaners which assist in producing a frequency and intensity stabilized beam with a spatially clean mode. The squeezer generates squeezed vacuum states which are reflected off of a filter cavity and injected into the antisymmetric port to provide broadband quantum noise reduction. The beamsplitter is shown with the high-reflective surface facing the antisymmetric port rather than the laser, unlike current detectors, to minimize loss in the signal extraction cavity, but careful analysis of thermal effects is needed before finalizing the design. (From the Cosmic Explorer Horizon Study\cite {Evans:2021gyd})}
  \label{fig:reference_design}
\end{figure}

{\bf Design:} The detectors of the Cosmic Explorer observatory concept are dual-recycled
Fabry--P\'{e}rot Michelson interferometers, the same as Advanced LIGO, scaled up to use 
\SI{40}{\km} or \SI{20}{\km} long arms. The longer arm
length will increase the amplitude of the observed signals with effectively no
increase in the noise.  The optical layout of the detectors is shown in figure \ref{fig:reference_design}, and the key design parameters are listed in table \ref{tab:detector_params}. Although there are areas of detector technology where
improvements will lead to increases in the sensitivity and bandwidth of the
instrument relative to the existing LIGO detectors, the dominant improvement
will come from the order-of-magnitude increase in length.

The interferometers installed in the Cosmic Explorer observatories will adapt
as the technologies and science evolve, and like LIGO and Advanced LIGO,
Cosmic Explorer's sensitivity is expected to improve with time due to technology upgrades
and commissioning effort.  Parts of the Cosmic Explorer nominal design may not be installed
before the Cosmic Explorer observatories begin collecting data.  These planned upgrades,
to be installed as the technology becomes available, may include: low-loss readout of high-fidelity squeezed states of light, adding seismometer arrays to
subtract fluctuations in the local gravity, and improved sensors for seismic
isolation relative to what is expected to be available at the time of
construction.

To minimize the required technical development, the initial CE detectors will use the Advanced LIGO detector design, including its A+ upgrades, scaled as needed in size, along with some advances to improve the low frequency sensitivity. This provides a straightforward approach to significant improvement using tested technology with relatively low risk. The planned upgrades will then proceed when possible given availability of new technologies and when maximally beneficial to the scientific output of the observatories. That is, the upgrades can all be performed in parallel at one or both observatories, or sequentially at one observatory at a time to avoid long down times.

\begin{figure}[!ht]
  \centering
  \includegraphics[width=0.8\textwidth]{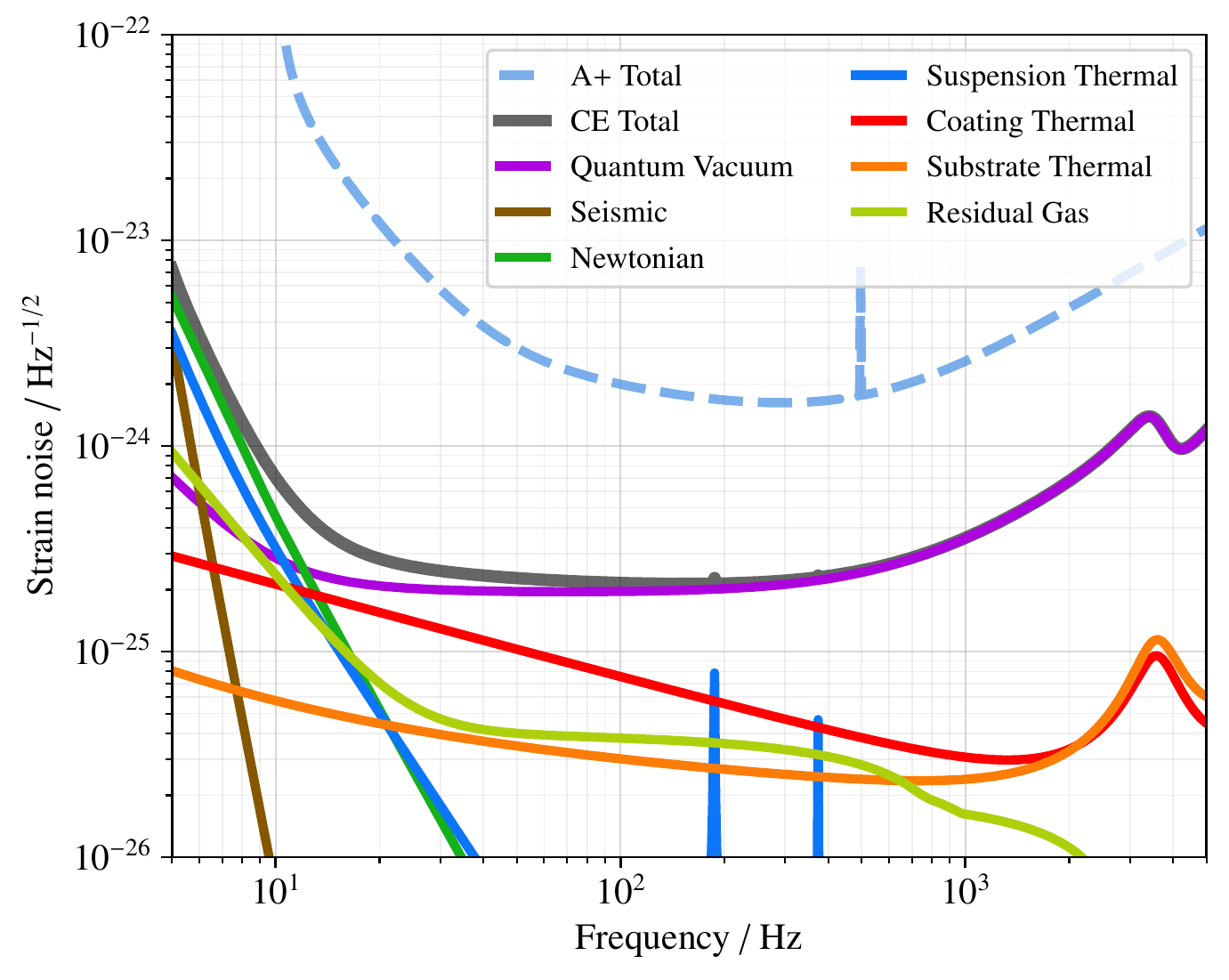}
  	\caption{Estimated spectral sensitivity (solid black) of Cosmic Explorer (CE) and the known fundamental sources of noise that contribute to this total (colored curves). The design sensitivity of LIGO A+ is also shown in dashed blue.  (From the Cosmic Explorer Horizon Study\cite {Evans:2021gyd})}
  	\label{fig:nb}
\end{figure}

{\bf Sensitivity:} The expected detector strain sensitivity of the Cosmic Explorer \SI{40}{\km} baseline design is shown in figure \ref{fig:nb}, together with an estimate of the ultimate performance of the A+ upgrade of the current Advanced LIGO detectors. The detector reaches a strain sensitivity of about \SI{2.5e-25}{\big/\!\sqrt{\Hz}} over a wide band.
Over much of that sensitivity band it will be limited by the residual optical quantum vacuum fluctuations (purple trace in figure \ref{fig:nb}), after they have been suppressed by about 10~dB from the regular quantum vacuum by frequency-dependent optical vacuum squeezing technology. This technology has already been demonstrated in Advanced LIGO and GEO600 \cite{PhysRevLett.123.231107,Vahlbruch:2010cy}, with GEO600 having achieved 6dB of effective squeezing of the interferometer readout.
The next dominant source of noise is coating Brownian thermal noise (red trace in figure \ref{fig:nb}). This thermal noise is driven by the mechanical dissipation in the optical coatings. The design assumes the same specifications for the coatings as the current A+ upgrade \cite{DoublingRange:2015} - with any additional improvement due to the larger optical beam and longer arm length. Despite this improvement, coating thermal noise still contributes significantly between about \SI{20}{\Hz} and \SI{100}{\Hz}. Research is underway to to find alternative coating materials, such as crystalline GaAs/AlGaAs \cite{Cole2013}. This coating could provide much lower coating thermal noise, but scaling it to the size of Cosmic Explorer optics requires further development.

Finally, most of the detector design changes are aimed at extending the observation band to lower frequencies, improving the seismic noise, Newtonian noise (i.e. direct gravitational coupling of seismic noise) and the suspension thermal noise (figure \ref{fig:nb}). Each of the four Cosmic Explorer test masses will be suspended by a quadruple pendulum to isolate them from seismic disturbances \cite{2012CQGra..29w5004A}. The suspensions provide passive $1/f^8$ filtering of seismic noise above their mechanical resonance frequencies. The suspensions themselves will be mounted on inertial seismic isolation systems which provide additional active and passive suppression of seismic noise \cite{2015CQGra..32r5003M}.
The inertial seismic isolation systems are similar to those of Advanced LIGO \cite{2015CQGra..32r5003M} but with improved inertial and position sensors. It is assumed that incremental improvements will allow Cosmic Explorer to initially achieve a threefold improvement over Advanced LIGO at \SI{10}{\Hz} and a tenfold improvement at \SI{1}{\Hz}. Novel six-dimensional inertial isolators with optical readout will be used to achieve an additional threefold improvement at \SI{10}{\Hz} and tenfold improvement at \SI{1}{\Hz} to achieve the final Cosmic Explorer sensitivity. 

{\bf Technology development:} Scaling the Advanced LIGO detector technology to Cosmic Explorer requires research targeting improvements in squeezing and quantum metrology techniques (see section \ref{sec:quantum_meas}), the production of large (320~kg) low-loss fused silica optics for test masses, optical coatings with reduced mechanical dissipation, and a low-cost ultra-high vacuum system. Accessing the scientifically interesting low-frequency band also requires improved active seismic isolation, including systems to subtract the direct Newtonian coupling of the seismic motion.

One possible upgrade for the Cosmic Explorer facilities is the technology currently being developed for the LIGO Voyager detector, consisting of a 2 micron laser and cryogenic silicon test masses. This approach would also require the production of large (320~kg) single crystal silicon test masses, a cryogenic cooling system with low vibrational coupling, and improved 2~um wavelength laser technology, particularly low-noise lasers and high quantum-efficiency photo diodes. The design parameters of this upgrade are also listed in table \ref{tab:detector_params}. Currently we anticipate that this cryogenic upgrade will only be required for Cosmic Explorer if a major problem is encountered with scaling up current technology.  

{\bf Cost:} The initial cost estimate for the Cosmic Explorer reference concept consisting of a 40~km detector and a 20~km detector is approximately $\$1.6B$ (2021~USD), as published in the Horizon Study \cite{Evans:2021gyd}.
It is based on extrapolating actual costs from LIGO construction, the Advanced LIGO upgrade,
and the work of professional civil engineering and metallurgy consultants.
The exercise of developing this cost estimate brought to the forefront a set
of cost-drivers which impact the technical design and scientific output
of a Cosmic Explorer observatory: arm length, beam tube material and diameter, and observatory location.
Notably, the cost of the detectors installed in the observatories is not a major driver.

The length of an observatory's arms is the most fundamental feature in
determining its potential scientific output.  Arm length needs to be increased to
the optimal length dictated by the science goals.
Many of the costs associated with arm length are simply proportional to the
length.  Examples of this are: the road which goes along the beamline and provides
access to the beamtube, the electrical utilities which run alongside the beamtube, the
slab which supports the beamtube, the beamtube enclosure, and the beamtube itself.  All
of these civil engineering costs are largely location independent (generally
within \SI{10}{\%} of the national average, and often a few percent lower than
the average for the reference sites considered for CE). 

The cost of excavation and transportation is not included in the above list of
civil engineering costs because it is highly location dependent, and generally
not proportional to the length of the facility.  As a concrete example,
consider a large dry lake bed (e.g., the Bonneville Salt Flats along interstate
80 west of Salt Lake City, UT).  The surface at such a location follows the
geoid almost perfectly: meaning that it follows the curvature of the Earth and
has constant altitude.  The arms of an observatory must, however, be straight lines
since laser beams do not curve with the Earth's geoid.  The curvature of the
Earth is such that the elevation at the center of a \SI{40}{\km} long straight
line is \SI{30}{\meter} lower than the ends.  Preparing such a site would require
excavating almost 10 million cubic meters of soil, and transporting it more
than \SI{10}{\km} on average (i.e., from the center to the ends), at a cost of
very roughly \$100 million (highly dependent on geology). This flat-site example drove significant interest in finding sites which minimize excavation
and transportation costs.
Such sites are slightly bowl-shaped with an elevation profile roughly \SI{30}{m} higher at
the ends than in the middle.
There are a number of wide ``valleys'' that fit this description in the western states,
and picking a location and orientation well can vastly reduce excavation
and transportation costs.
%However, this search for topographically favorable sites clearly showed that the number of advantageous and available sites decreases rapidly with arm length, meaning that excavation costs for a \SI{20}{\km} observatory may be less than a \SI{5}{\%} of the total observatory cost, while for a \SI{40}{\km} observatory they are likely to remain near \SI{10}{\%} of the total simply because there are fewer \SI{40}{\km} sites to choose from.

Finally, as with all gravitational-wave detectors, ultrahigh vacuum (UHV) is necessary
in Cosmic Explorer to reduce path-length fluctuation of the light traveling down the arms and
to reduce mechanical damping on the detector’s core optics. The diameter of this vacuum system
is set by the back-scatter requirements from the optical beam, requiring a diameter of several
tens of centimeters along its entire length. The material and construction cost of this vacuum
system is a major cost driver for Cosmic Explorer, and research to mitigate this cost without compromising
the vacuum quality is underway.

\begin{table}[t]
    %\sisetup{
    %}
    %\centering
    %\sisetup{
    %table-format=3.1,
    %table-space-text-post=\hspace{-1em}
    %}
    \begin{tabular}{r l S S S S}
    \toprule
         &
         {\textsf{Quantity}}      &
         {\textsf{Units}}   &
            {\textsf{LIGO A+}}       &
            {\textsf{CE}} &
            {\textsf{CE (\SI{2}{\um})}}    \\
    \midrule
        &
            Arm length  &
            \si{\km} &
            4 & 40 & 40 \\
        &
            Laser wavelength  & \si{\um} &
            1 & 1 & 2 \\
        &
            Arm power  & \si{\MW} &
            0.8 & 1.5 & 3 \\
        &
            Squeezed light & \si{\dB} &
            6 & 10 & 10 \\
        &
            Susp. point at \SI{1}{\Hz} & $\si{\pico\meter\big/\sqrt{\text{Hz}}}$ &
            10 & 0.1 & 0.1 \\

        \midrule
        Test masses &
            Material & &
            {Silica} & {Silica} & {Silicon} \\
        &
            Mass & \si{\kg} &
            40 & 320 & 320 \\
        &
            Temperature & \si{\K} &
            293 & 293 & 123 \\

        \midrule
        Suspensions &
            Total length & \si{\meter} &
            1.6 & 4 & 4 \\
        &
            Total mass & \si{\kg} &
            120 & 1500 & 1500 \\
        &
            Final stage blade & &
            {No} & {Yes} & {Yes} \\

        \midrule
        Newtonian noise &
            Rayleigh wave suppr. & \si{\dB} &
            0 & 20 & 20 \\
        &
            Body wave suppr. & \si{\dB} &
            0 & 10 & 10 \\

        \midrule
        Optical loss &
            Arm cavity (round trip) & \si{\ppm} &
            75 & 40 & 40 \\
        &
            SEC (round trip) & \si{ppm} &
            5000 & 500 & 500 \\

    \midrule
    \midrule
        & BNS horizon redshift & & 
            \aplusBnsHorizon{}        &
            \ceTwoGlassBnsHorizon{}   &
            \ceTwoSiliconBnsHorizon{} \\
        & BBH horizon redshift & &
            \aplusBbhHorizon{}        &
            \ceTwoGlassBbhHorizon{}   &
            \ceTwoSiliconBbhHorizon{} \\
        & BNS SNR, $z=0.01$  & &
            \aplusBnsSNR{}        &
            \ceTwoGlassBnsSNR{}   &
            \ceTwoSiliconBnsSNR{} \\
        & BNS warning, $z=0.01$  & \si{\minute} & 
            \aplusEarlyWarning{}        &
            \ceTwoGlassEarlyWarning{}   &
            \ceTwoSiliconEarlyWarning{} \\
    \bottomrule
    \end{tabular}
    \caption{
        Key design parameters and astrophysical performance measures for the
LIGO A+ and \SI{40}{\km} Cosmic Explorer detectors.  The astrophysical performance measures assume a 1.4--1.4$M_\odot$ binary-neutron-star (BNS) system and a
30--30$M_\odot$ binary-black-hole (BBH) system, both optimally oriented.
``BNS warning'' is the time before merger at which the event has accumulated
a signal-to-noise ratio (SNR) of 8. Acronym: SEC stands for signal extraction cavity.
    \label{tab:detector_params}}
\end{table}

\subsection{Organization and Schedule of Cosmic Explorer}

\begin{figure}[t]
    \centering
    \includegraphics[width=\textwidth]{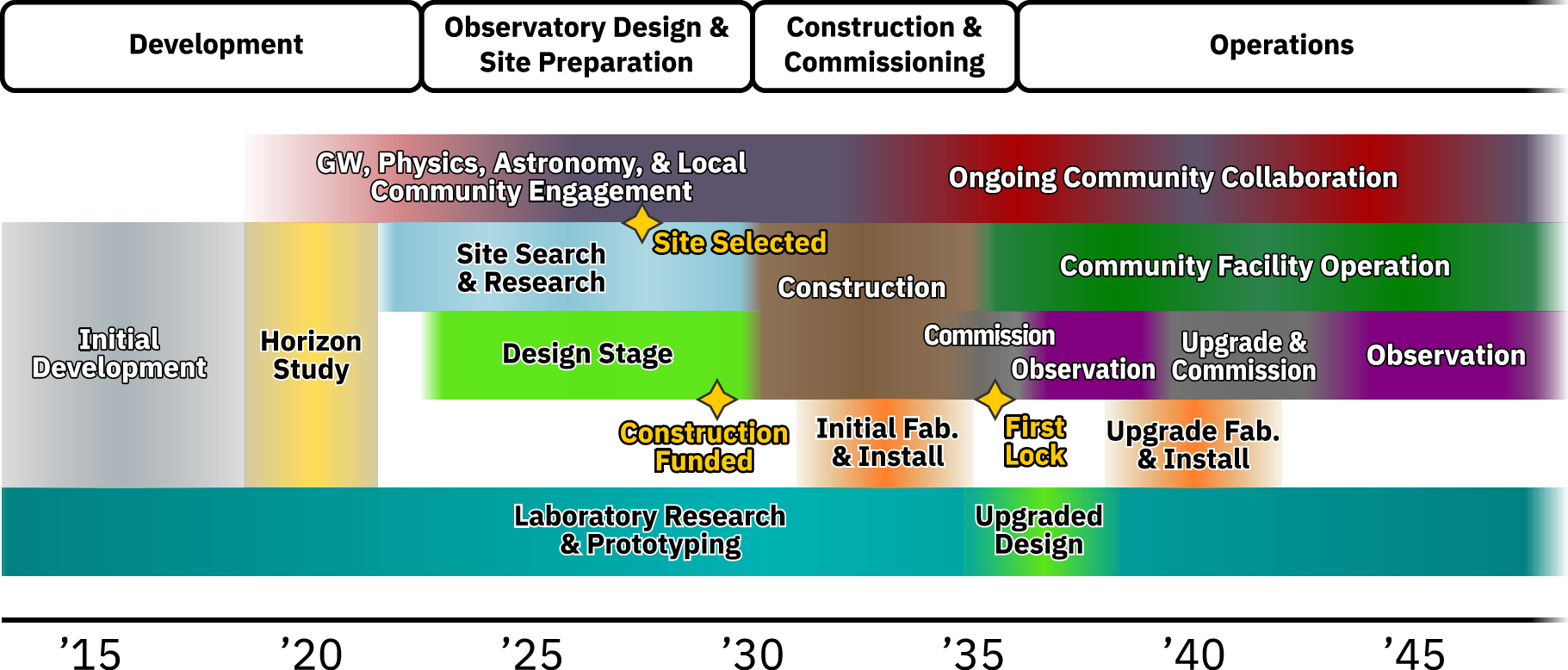}
    \caption{Cosmic Explorer top-level timeline showing a phased approach to design and construction. The eventual divestment from the facility is not indicated. (From \cite{Evans:2021gyd}.)}
    \label{timeline}
\end{figure}

The envisioned timeline for Cosmic Explorer spans multiple decades and takes place in distinct stages:
development; observatory design and site preparation;
construction and commissioning; initial operations;
planned upgrades; operations at nominal sensitivity;
future observatory upgrades and operations.
The development stage for Cosmic Explorer began in 2013, and has culminated in the publication of the Cosmic Explorer Horizon Study \cite{Evans:2021gyd}. It will be followed by the design and site selection stage, targeting a start of construction in 2029, and initial observations in 2035. Figure \ref{timeline} graphically summarizes this timeline.

\subsection{Voyager Upgrade to Current Facilities}
%[Rana / Chris Wipf] 
\label{sec:voy}

The current LIGO detectors will approach the thermodynamic and quantum
mechanical limits of their designs within a few years. Over the next
several years, aLIGO will undergo a modest upgrade, designated
``A+''. The aim of this upgrade is chiefly to lower the quantum (shot)
noise through the use of squeezed light, and also to reduce somewhat
the thermal noise from the mirror coatings. This upgrade has the goal of enhancing the sensitivity by
$\sim$50\% \cite{DoublingRange:2015}.

% -- removed this; it seems to astro-y and not enough physics-y
%GW observations of compact binary systems provide a great deal of information about the component objects as well as the populations of NSs and BHs in the uiverse. The analysis of individual detections yields the masses and spins of the compact objects involved \cite{PhysRevLett.116.241102, PhysRevD.91.042003, PhysRevD.95.064052}. The distribution of these parameters in the population, along with the overall merger rates, will give critical insights into the processes that govern binary evolution. These include mass transfer in progenitors of compact binaries, supernova kicks, the efficiency of common-envelope ejection, and the dominant processes governing dynamical binary formation in dense stellar environments (e.g. Mandel\cite{Mandel_2010}, Postnov\cite{Postnov:2014}, Rodriguez\cite{PhysRevD.93.084029}, and references therein).

Voyager\cite{Adhikari_2020} represents a more substantial upgrade that can increase the range by a factor of 4\,--\,5 over
Advanced LIGO, and the event rate by approximately 100 times. Such a dramatic change in the sensitivity should
increase the detection rate of binary neutron star mergers to about
10 per day and the rate of binary black hole mergers to around 30 per day.
This upgraded instrument would be able to detect binary black holes
out to a redshift of 8.

The path to Voyager requires aggressively reducing several noise sources, including: quantum radiation pressure and shot noise, mirror thermal noise, mirror suspension thermal noise, and Newtonian gravity noise. All of these noise sources are addressed by the Voyager design, with the goal of commissioning and observational runs in about a decade. The most significant design changes can be traced to the need to reduce the quantum noise in tandem with the mirror thermal noise.

\begin{itemize}
\item Quantum noise will be reduced by increasing the optical power stored in the interferometers.
In Advanced LIGO, the stored power is limited by thermally induced wavefront distortion effects in the fused silica test masses.  These effects will be alleviated by choosing a test mass material with a high thermal conductivity, such as silicon. 

\item The test mass temperature will be lowered to 123\,K, to mitigate thermo-elastic noise.  This species of thermal noise is especially problematic in test masses that are good thermal conductors.  Fortunately, in silicon at 123\,K, the thermal expansion coefficient crosses zero, which eliminates thermo-elastic noise. Other plausible material candidates, such as sapphire, require cooling to near 20\,K to sufficiently suppress this noise.

\item The thermal noise of the mirror coating will be reduced by switching to low dissipation amorphous silicon based coatings, and by reducing the temperature.
\end{itemize}

Put together this cryogenic interferometer design exploits the full physical limits of the existing LIGO facilities, and will enable a qualitatively brighter vision of the dark side of the universe.

\subsection{International Partner Projects}

Several of the main science goals of next-generation gravitational-wave detectors depend on good localization of sources on the sky, requiring a global network of detectors. International collaboration is essential to establish such a network and operate it concurrently. Currently Europe and Australia have active communities developing plans for such detectors, specifically:

{\it Einstein Telescope (ET)}: 
The Einstein Telescope \cite{Punturo:2010zz} is a planned next-generation gravitational-wave observatory in Europe. It consists of an underground triangular facility with 10~km arm length, housing multiple interferometers. The underground location suppresses the expected seismic disturbances from surface Rayleigh waves, reducing the Newtonian noise that limits all ground-based gravitational-wave facilities a low frequencies. The sensitivity at those low frequencies permits the observation of heavier compact binary mergers and increases the observatory's early-warning time substantially.

The Einstein Telescope has been added to the road map of the European Strategy Forum on Research Infrastructure (ESFRI). The road map identifies the most promising European scientific structures based on an in-depth evaluation and selection procedure.

Given the similar size and scope of Einstein Telescope and Cosmic Explorer, as well as the need for a global observatory network, technical collaborations between the two projects is desirable and ongoing. For example, the development of the vacuum system technology is critical to both projects, and a research collaboration including the CERN vacuum group and both Cosmic Explorer and Einstein Telescope is underway.

{\it Neutron star Extreme Matter Observatory (NEMO)}:
The Neutron star Extreme Matter Observatory~\cite{Ackley:2020atn} is focused on observing the late in-spiral and post-merger signatures of Binary Neutron Star mergers, to obtain their equation of state~\cite{Bernuzzi2020}. NEMO in a laser interferometric gravitational wave detector with a target strain sensitivity reaching at least \SI{1e-24}{\big/\!\sqrt{\Hz}} in a window around $2\,$~kHz. In that window the sensitivity would be of a comparable sensitivity to Cosmic Explorer \cite{Evans:2021gyd} and the Einstein Telescope \cite{Punturo:2010zz}, but would be achieved with a specific detector configuration that targets high frequencies.

The central design philosophy is to considerably relax the sensitivity requirements at low-frequencies (below $1\,$~kHz) and to enable high bandwidth control loops to mitigate opto-mechanical instabilities~\cite{Evans:2015} and reduce the cost with less complex test mass suspension systems. The optical configuration will be similar to the current observatories using a dual-recycled Fabry-Perot Michelson interferometer. Compared to other detectors however, it will employ a long signal recycling cavity to tune and maximise its sensitivity in the 1–4 kHz band. To further improve the target sensitivity, alternative signal enhancement techniques can potentially be utilised~\cite{Adya2020,Page2021.Communications-Physics}.  NEMO is envisioned to employ third generation technologies, such as the use of longer laser wavelengths~\cite{Kapasi2020.OE} and cryogenically cooled silicon test masses~\cite{PhysRevD.102.122003}. This can make a NEMO detector a pathfinder for technologies that will enhance the next-generation detectors.

\section{Atom Interferometers}
\label{sec:ai}
%\swb{Should this go to Section 5, intro?}
In the past several years, there has been widespread and growing international interest in pursuing long-baseline atomic sensors for gravitational wave detection and ultralight wave-like dark matter searches.  An impressive number of efforts have begun around the world, including both terrestrial experiments and space-based proposals. In the US, MAGIS-100\cite{Abe_2021} is an intermediate-size detector with a 100-meter baseline currently under construction at Fermilab. In Europe, significant progress has already been made on the construction of MIGA (Matter wave-laser based Interferometer Gravitation Antenna)\cite{canuel2018MIGA}, a $200~\text{m}$ baseline underground gravitational wave detector demonstrator located in France.  To follow up on this, a new proposal has called for the construction of ELGAR (European Laboratory for Gravitation and Atom-interferometric Research) \cite{Canuel2019ELGAR}, an underground detector with horizontal $32~\text{km}$ arm length aiming to detect gravitational waves in the mid-band (infrasound) frequency range.  In China, work has begun to build ZAIGA (Zhaoshan long-baseline Atom Interferometer Gravitation Antenna) \cite{ZAIGA2020}, a set of multiple $300~\text{m}$ vertical shafts separated by km-scale laser links that will use atomic clocks and atom interferometry to explore a wide range of science including gravitational wave detection.  In the UK, a broad collaboration of eight institutes has recently advanced a multi-stage program called AION (Atom Interferometer Observatory and Network) \cite{Badurina_2020}, which aims to progressively construct atom interferometers at the 10- and then 100-meter scale, in order to develop technologies for a full-scale kilometer baseline instrument for both gravitational wave detection and dark matter searches.  To access lower frequencies, a variety of space-based detectors have also been proposed, based both on atomic clocks \cite{kolkowitz2016gravitational,ebisuzaki2020ino} and atom interferometers \cite{graham2017mid,hogan2019gravitational,abou2020aedge,hogan2011atomic}, and in fact these technologies are closely related \cite{norcia2017role}.

The ambitious scope of these experiments and proposals is evidence of the enthusiasm in the community for the long-term science prospects offered by long-baseline atomic sensing.  In addition, these many detectors have the potential to complement each other. The diversity of approaches taken by the various experiments is a clear strength, offering opportunities for different groups to develop alternate atomic sensing technologies in parallel.  More directly, operating multiple detectors in different parts of the world as part of a network offers valuable scientific advantages \cite{Badurina_2020}.  In the spirit of the LIGO/Virgo/KAGRA collaboration, correlating data collected simultaneously by several atom interferometer gravitational wave detectors operating in the mid-band frequency range would be a powerful way to improve background rejection and increase overall sensitivity.

\subsection{Technology Description and Status}

Atom interferometers offer a promising route to gravitational wave detection in the `mid-band' frequency range between $0.03~\text{Hz}$ and $3~\text{Hz}$ with terrestrial and space-borne instruments.  The detection concept~\cite{graham2013new,graham2017mid} takes advantage of features of both atom interferometers and atomic clocks to allow for a single-baseline gravitational wave detector~\cite{PhysRevD.78.042003,Yu2011,graham2013new}.  Additionally, the same detector configuration enables the exploration of new regions of dark-sector parameter-space~\cite{BRNreport} by being sensitive to proposed scalar- and vector-coupled dark matter candidates in the ultralight range ($10^{-15}$~eV -- $10^{-14}$~eV).

Dilute clouds of freely-falling ultra-cold Sr atoms simultaneously serve as inertial references and as precise clocks. Laser light propagates between two atom ensembles separated by a baseline. The laser pulses drive transitions between the ground and excited states of the Sr clock transition (${}^1S_0 \rightarrow {}^3P_0$), the same line used in state-of-the-art optical lattice clocks~\cite{campbell2017fermi}. A sequence of short light pulses generates a pair of atom interferometers, one on each end of the baseline~\cite{Hogan2009}.  Specifically, an initial pulse splits the atom into a quantum superposition of the ground and excited clock states, corresponding to the two arms of the atom interferometer.  Excitation of atom corresponds to the absorption of a photon (including both the photon's energy and its momentum), and the portion of the atomic wave function in the excited state therefore also receives a single photon momentum recoil.  The momentum difference between the two arms of the interferometer causes these arms to separate spatially after free evolution.  Additional pulses further split, recombine, and interfere the two interferometer arms while toggling each arm between the ground and excited clock states.  
%The timing of the atomic transitions, and thus the time the atoms spend in a superposition of the ground and excited states, depends on the change of the light travel time across the baseline, $L/c(\dot{h}\Delta T)$~\cite{graham2013new}. The resulting atom interferometer phase $\phi$ is then proportional to the length $L$ of the baseline: $\phi \,\propto\,\omega_A L/c$, where $\hbar\, \omega_A$ is the energy splitting of the clock transition. \swb{Make it more explicit for outsiders to follow--Have attempted to make the discussion above more explicit.} As a result, the differential phase measurement between the two atom interferometers is sensitive to variations in both the baseline $L$ and the clock frequency $\omega_A$ that arise during the light-pulse sequence. 
The timing of the atomic transitions of the far atom interferometer, and thus the time the atoms spend in a superposition of the ground and excited states, depends on the change of the light travel time across the baseline, $\Delta L/c$~\cite{graham2013new}, where $\Delta L = L(\dot{h} \Delta T)$ is the effective change in arm length, $h$ is the gravitational wave strain and $\Delta T$ is the time between laser pulses.
As a result, the differential phase measurement between the two atom interferometers is proportional to $\Delta \phi \,\propto\,\omega_A \Delta L/c$, and thus is sensitive to variations in both the baseline $\Delta L$ and the clock frequency $\omega_A$ that arise during the light-pulse sequence. 
A passing gravitational wave modulates the baseline length, while coupling to an ultralight dark matter field can cause a modulation in the clock frequency.  Thus, this measurement concept combines the prospects for both gravitational wave detection and dark matter searches into a single detector design, and both science signals are measured concurrently.

The sensitivity of atom interferometers to gravitational waves is driven by several key parameters.  First, the sensitivity scales proportionally to the baseline length $L$.  Sensitivity can also be enhanced through the use of advanced large momentum transfer (LMT) atom optics which deliver a large number $n$ of photon momentum kicks to the atoms, as well as by reducing phase noise in the matter-wave interference fringes (atom shot noise) through the use of both high flux atom sources and quantum entangled atoms.  Much of the current interest in exploring atom interferometric gravitational wave detectors can be attributed to rapid improvements in these areas.  Atom interferometers with a 10~m height have been realized with LMT of $n \sim 100$ \cite{kovachy2015quantum}, and entangled ensembles of atoms have been demonstrated with phase noise variance 100 times below the standard quantum limit \cite{hosten2016measurement}.  It is important to note that atom interferometry using the Sr clock transition~\cite{hu2017atom} offers the potential for multiple order of magnitude increases in $n$ due to the low associated spontaneous emission rate \cite{Rudolph2020}.  Current research and development efforts aim to continue these advances, as summarized in Table \ref{table:futurevision} for the MAGIS research program.

\begin{table}
\centering
\footnotesize
\caption{Detector design parameter targets for the MAGIS-100 detector at Fermi National Accelerator Laboratory and envisioned follow-on detectors.  The baseline length $L$ is the total end-to-end length of the detector. LMT atom optics of order $n$ refers to an $n\hbar k$ momentum splitting between the two arms of the atom interferometer (corresponding to $n$ photon recoil kicks).  The atom phase noise $\delta\phi$ listed is for a single atom source, and assumes improvements in atom flux and the use of spin-squeezed atomic states~\cite{hosten2016measurement,PhysRevLett.116.093602}. The multiple atom sources are assumed to be distributed uniformly along the baseline. MAGIS-100 (initial) corresponds to current state-of-the-art parameters, while (final) assumes atom optics operating at the projected physical limit for this baseline. The space-based configuration is discussed in greater detail in~\cite{graham2017mid}.
}
\label{table:futurevision}
\begin{tabular}{lccccc} 
 \toprule
   Experiment & (Proposed) Site & Baseline & LMT Atom   & Atom    &  Phase Noise  \tabularnewline
  &   &  $L$ (m)  & Optics $n$ & Sources & $\delta\phi$ ($\text{rad}/\sqrt{\text{Hz}}$)  \tabularnewline
  \midrule
 Sr prototype tower & Stanford & $10$ & $10^2$ & 2 & $10^{-3}$ \tabularnewline
 MAGIS-100 (initial) & Fermilab (MINOS shaft) & $100$ & $10^2$ & 3 & $10^{-3}$ \tabularnewline
 MAGIS-100 (final) & Fermilab (MINOS shaft) & $100$ & $4\times10^4$ & 3 & $10^{-5}$ \tabularnewline
 MAGIS-km & Homestake mine (SURF) & $2000$ & $4\times10^4$ & 40 & $10^{-5}$ \tabularnewline 
 MAGIS-Space & Medium Earth orbit (MEO) & $4\times 10^7$ & $10^3$ & 2 & $10^{-4}$ \tabularnewline
 \bottomrule
\end{tabular}
\end{table}

Projected gravitational wave strain sensitivity curves for the detectors described in Table \ref{table:futurevision}, including the MAGIS-100 detector currently under construction at Fermilab and follow-on detectors, are shown in Fig. \ref{fig:MAGIS-gw-sensitivity-100}.  These curves assume detectors limited by atom quantum projection noise in the frequency band shown, requiring the detectors to be designed so that technical noise sources are below quantum projection noise in this frequency band \cite{Abe_2021}.  The continued development of techniques to suppress various technical noise sources in the detector is therefore essential \cite{Abe_2021}.  Key recent advances include the implementation of spatially resolved detection to mitigate noise backgrounds coupled to initial atom cloud kinematics \cite{Dickerson2013,Sugarbaker2013}, the development of rotation compensation protocols to counteract Coriolis forces from the Earth's rotation \cite{Lan2012,Dickerson2013,Sugarbaker2013,asenbaum2020atom}, and the realization of atom clouds with effective temperatures as low as 50~pK to reduce backgrounds associated with cloud expansion (such as the coupling of cloud expansion to laser wavefront imperfections)~\cite{kovachy2015matter,Rudolph2016Thesis}.  Newtonian noise, which is indicated separately in Fig. \ref{fig:MAGIS-gw-sensitivity-100}, is another important noise background.  Measurement protocols are being investigated to characterize and suppress the influence of Newtonian noise \cite{PhysRevD.93.021101, Abe_2021}.  Since Newtonian noise is an important background for all terrestrial detectors \cite{HarmsGGN}, this research can benefit both the laser interferometer and atom interferometer approaches to gravitational wave detection.
%\swb{Highlight that this problem is common to all terrestrial detectors, and the same research can benefit both approaches to GW detection--Have reworded to emphasize this point}

\subsection{MAGIS and AION}

\begin{figure}[t]
\begin{center}
\includegraphics[width=1\textwidth]{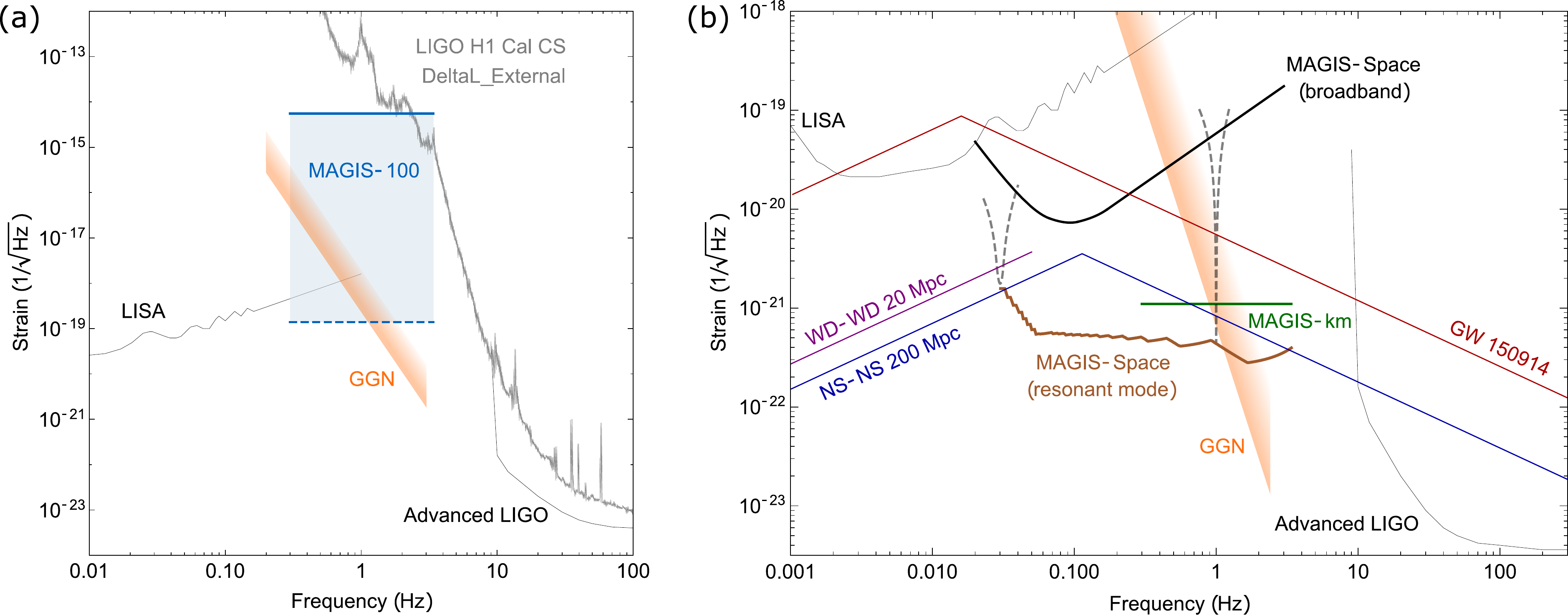}
\end{center}
    \caption{(a) Projected gravitational wave strain sensitivity for MAGIS-100 and follow-on detectors. The solid blue line shows initial performance using current state of the art parameters (table \ref{table:futurevision}, initial). The dashed line assumes parameters improved to their physical limits (table \ref{table:futurevision}, final). LIGO low frequency calibration data (gray) is shown as an estimate for the state-of-the-art performance in the mid-band frequency range~\cite{PhysRevD.102.062003}. An estimate of Newtonian noise (gravity gradient noise, GGN) at the Fermilab site is shown as an orange band. (b) Estimated sensitivity of a future km-scale terrestrial detector (MAGIS-km, green) and satellite-based detector (MAGIS-Space, brown) using detector parameters from table \ref{table:futurevision}.  The detector can be switched between both broadband (black, solid) and narrow resonant modes (black, dashed).  The resonant enhancement $Q$ can be tuned by adjusting the pulse sequence~\cite{Graham:2016plp}. Two example resonant responses are shown targeting $0.03~\text{Hz}$ ($8\hbar k$ atom optics, $Q=9$) and $1~\text{Hz}$ ($1\hbar k$ atom optics, $Q=300$).  The brown curve is the envelope of the peak resonant responses, as could be reached by scanning the target frequency across the band.  Sensitivity curves for LIGO~\cite{AdvancedLIGO2015} and LISA~\cite{prince2002lisa} are shown for reference. Also shown are a selection of mid-band sources including neutron star (NS) and white dwarf (WD) binaries (blue and purple) as well as a black hole binary already detected by LIGO (red). The GGN band (orange) is a rough estimate based on seismic measurements at the SURF site~\cite{Harms_2010}.}
\label{fig:MAGIS-gw-sensitivity-100}
\end{figure}

The MAGIS research program aims to develop a family of sensors based on the concept outlined above with increasing baseline length and sensitivity, as summarized in Table \ref{table:futurevision}.  MAGIS-100 is the first detector is the first detector in this family \cite{Abe_2021}. The instrument features a 100-meter vertical baseline and is now under construction at the Fermi National Accelerator Laboratory (Fermilab). State-of-the-art atom interferometers are currently operating at the 10-meter scale~\cite{Dickerson2013,kovachy2015quantum,asenbaum2016phase,asenbaum2020atom,Hartwig2015,zhou2011development}, while a kilometer-scale detector is likely required to detect gravitational waves from known sources. MAGIS-100 is the first step to push the limits of atom interferometry beyond the lab-scale and bridge the gap to future detectors. It is designed to be operated in the manner of a full-scale detector and aims to achieve the high up-time required from such a facility. MAGIS-100 will explore a wide variety of systematic errors and backgrounds to serve as a technology demonstrator for future gravitational wave detection with atom interferometry. Additionally, the detector is expected to be sensitive enough to search for potential ultralight dark matter couplings beyond current limits.  By operating in two distinct dark matter search modes, MAGIS-100 can look for both scalar-coupled and vector-coupled dark matter candidates in so-far unexcluded regions of parameter space. Finally, by extending the scale of matter-wave interferometers to a 100-meter baseline, MAGIS-100 has the opportunity to advance the frontier of quantum science and sensor technologies, including tests of the validity of quantum mechanics in a regime in which massive particles are delocalized over record-setting macroscopic time~\cite{Dickerson2013,xu2019probing} and length~\cite{kovachy2015quantum,asenbaum2016phase} scales.  Projected gravitational wave strain sensitivity curves for MAGIS-100 and follow-on detectors are shown in Fig. \ref{fig:MAGIS-gw-sensitivity-100}.

AION (Atom Interferometer Observatory and Network)~\cite{Badurina_2020} is UK-based multi-stage program, which aims to progressively construct atom interferometers at the 10- and then 100-meter scale, in order to develop technologies for a full-scale kilometer-baseline instrument for both gravitational wave detection and dark matter searches.  It will exploit the same detector concept as MAGIS, and MAGIS and AION are collaborating closely on detector design and research and development efforts.  A substantial focus of the AION research program is studying how to leverage a global network of atom interferometric gravitational wave and dark matter observatories~\cite{Badurina_2020}.  These terrestrial instruments may also ultimately pave the way for space-based detectors with even greater sensitivity.  In anticipation of this, initial studies of the requirements of and scientific motivation for space-based, long-baseline atom interferometers have already begun~\cite{graham2017mid,loriani2019atomic,abou2020aedge}.

\subsection{Future Atom Interferometers}

In the US, the MAGIS-100 experiment under construction will serve as a prototype detector to probe the key technologies required to build a Km-scale detector capable of detecting gravitational waves in the mid frequency band. The path from 100m to 1Km and eventually to space is a long-term program that will require broad international support and National Laboratory facilities, resources, and expertise for the construction and operation of large-scale experiments. Key areas of R\&D focus on two main areas: large-scale challenges, and advancing quantum sensing technologies. A DOE involvement in this technology could over the next 10 years lead to an international collaboration similar to the LIGO-Vigo-KAGRA Scientific Collaboration. 

DOE investment in long-baseline atom interferometers is (a) an opportunity to expand the physics program and technical knowledge of the DOE HEP program, (b) a necessity for these emerging large and complex experiments to be built on a foundation of DOE laboratory support and expertise, and (c) synergistic both with other DOE HEP science programs and more broadly with NSF.

\section{Common Technologies and Research Opportunities}
\label{sec:ct}

%DOE National Lab's role in gravitational-wave physics

%Check out CMB is Snowmass 2015 reports (Rana)

%Emphasize specific tech examples and DOE labs related capabilities https://drive.google.com/file/d/1D2bSnmpQ2h3Mn2e9tUSJDUKNW6BA6vfe/view

%Land use issues

%Examples: vacuum, Newtonian noise cancellation, etc.

%Gravitational wave experiments offer new and powerful ways to expand the exploration of the universe and probe fundamental physics. 
The gravitational-wave physics community is moving towards very large, extremely complex, facilities that will require addressing key technological and engineering challenges. DOE National laboratories could play a major role in enabling these projects capitalizing on the existing expertise in building and operating large-scale facilities such as particle accelerators and HEP detectors. For example, technical expertise related to accelerator design, such as ultra-high vacuum and magnetic field engineering, diagnostics and controls, as well as operational experience coordinating large-scale international projects, is widely applicable to the proposed future gravitational-wave  facilities. 

Potential synergies with DOE national labs and particle physics expertise include the development of larger UHV vacuum systems, active alignment systems for large precision facilities, new materials for mirror coating that can reduce the coating thermal noise, and low-temperature science with high cooling rates and low vibration such as low-vibration cyro-coolers, low-vibration heat transport and low temperature vibration sensors. In Europe, such collaborations have already begun with joint research of the CERN vacuum group with the Einstein Telescope, as well as the Cosmic Explorer project.
In the longer term, long-baseline atomic experiments promise to enable the exploration of the mid-band range of the gravitational-wave spectrum. This frequency range will allow observation of white dwarf binary mergers, heavier black holes mergers, and cosmological signals such as those from the electroweak phase transition or cosmic strings. Scaling up existing atom sensing technology to very long baselines will benefit from particle physics and accelerator physics expertise in the construction, alignment, monitoring, operation, and data analysis of large, complex, instruments.  
Gravitational-wave facilities present an unique new scientific and technological opportunity for DOE to expand its reach and scope of fundamental physics research. The science enabled by the next-generation of gravitational-wave facilities may lead to transformative advances in our understanding of the universe, complementing and extending the capabilities of future particle colliders and cosmology experiments, and enabling completely new directions to explore the early universe. 

A DOE involvement in third-generation gravitational-wave laser and atomic interferometers in partnership with NSF would leverage existing expertise at National Laboratories and  would be synergistic with other DOE fundamental science goals. 

\subsection{Large-scale Challenges}

A common theme for all gravitational-wave detector concepts is the need to scale up a detector from small proof-of-concept prototypes to large facilities. It is a direct consequence of gravity’s minimal coupling to matter and of the wavelength of astrophysical signals. The 20~and 40~km arm lengths of the Cosmic Explorer interferometers require dedicated new facilities. The size will drive the total project cost. This type of infrastructure has traditionally been within the purview of DOE National laboratories. Building these facilities also requires a large continuous stretch of land and will have a significant impact on the landscape, environment and the local community. As such the local community needs to be involved in the development, and an assessment of the environmental impact must be included in the site selection process from the beginning. 

The choice of site location for future detector facilities will be important for realizing the best possible detector performance. If a suitable shared site is identified, it may be desirable for atom and laser interferometers to leverage shared infrastructure. The correlation of data between atom interferometers, laser interferometers, and supplementary seismometer networks~\cite{HarmsGGN} may also prove to be valuable.
%Beyond site considerations, both atom and laser interferometers benefit from ultra-high-quality optics that minimize wavefront errors.

Scaling-up detectors to multiple-km-scale facilities will require National Lab core competencies in ultra high vacuum, mechanical, optical, and electrical engineering, precision alignment, magnetometry, high power lasers, diagnostics and monitoring, computing, distributed analysis infrastructure, and data analysis of large data sets. A Km-scale atom interferometer detector will demand large production of atom sources (around 20) which will require significant AMO expertise and resources that are beyond what is possible at universities. Existing expertise in particle accelerators, lasers, and large particle physics and cosmology experiments can be directly applicable to address these challenges.

\subsection{Newtonian Noise}
\label{sec:NN}

Fluctuations in the local gravitational force and direction due to seismic and atmospheric motion in the environment, known
as “Newtonian noise” or "gravity gradient noise", are the most significant fundamental low-frequency noise source for all terrestrial gravitational-wave detectors below a few Hertz. They are the dominant source of noise after the seismic contributions have been suppressed. Newtonian noise affects all detector designs equally, making it a research area of common interest. Similarly the detectors are also affected by Newtonian noise due to density fluctuations in the atmosphere at infrasonic frequencies. 

Because there is no way to shield or screen gravity, Newtonian noise difficult to directly suppress. Mitigation strategies include subterranean end- and corner stations to suppress the dominant seismic surface waves (Rayleigh waves), subtraction techniques that rely on precisely measuring the local seismic field using an array of seismometers, or even modifying the density of the material, intentionally deflecting or dissipating seismic waves with engineered materials or seismic metamaterials.

\subsection{Quantum Measurement}
% (Lee McCuller)
\label{sec:quantum_meas}

%QND, non-Gaussian, GKP, teleportation

Quantum sensing techniques are already at the heart of existing gravitational-wave detectors. Controlling the quantum vacuum in the optical modes in order to suppress the quantum shot- and radiation-pressure noise is done on a routine basis in the operating Advanced LIGO detectors. Much of the recent sensitivity improvements in these detectors is in fact due to this squeezed quantum vacuum technology, which was retrofitted in the detectors for the O3 observation run.
The use of squeezed vacuum in gravitational-wave interferometers leads to entanglement of the photons in the interferometer arms, which in turn imprints these fluctuations on to the interferometer mirrors through radiation pressure, entangling these macroscopic mirrors. The resulting quantum correlations between Advanced LIGO's 40~kg mirrors have been experimentally confirmed \cite{LIGOScientific:2020luc}. Scaled to the next-generation detectors, Cosmic Explorer will thus be using entangled 320~kg optics, separated by 35~miles.

Further improvements in the optical quantum measurement techniques are planned for the A+ upgrade, simultaneously suppressing radiation pressure and shot noise. All future detector concepts (Cosmic Explorer, Einstein Telecope, Voyager and NEMO) rely on further improvements in the sensing scheme. Most critically, this requires a tight control on optical losses and modes to avoid regular noisy quantum vacuum to leak into the readout sensing chain.

Similarly, for atom interferometers, achieving the ambitious scientific goals of Km-scale detectors will require expanding and advancing quantum information science competences at National Laboratories, leveraging the existing investment in the QuantiSED program, and QIS centers. To reach the necessary level of sensitivity, these detectors will leverage quantum states in which individual atoms will be coherently delocalized over record-setting and truly macroscopic distances of many meters. To achieve and optimally exploit these states, significant R\&D will be required in advanced quantum sensing techniques such as spin squeezing and increasing large momentum transfer. One example is quantum optimal control R\&D being carried out at DOE QIS centers which could lead to atom optics pulses with higher fidelity and better robustness to noise, helping the efforts to achieve larger momentum transfer atom optics.  These efforts aim to adapt quantum optimal control techniques used for microwave pulses that perform logical gates for superconducting qubits, an area of expertise at DOE National Laboratories, to optical pulses that manipulate the quantum states of atoms in an interferometer. 

Spin squeezing R\&D is aimed at enabling measurement of the atom interferometer phase with resolution beyond the standard quantum limit.  The incorporation of spin squeezing will lead to novel quantum states in which entanglement is combined with coherent delocalization over macroscopic distances, resulting in squeezing of the phase between the two macroscpically separated arms of the interferometer.  Other important R\&D items include:
i) Improving cooling of atoms by developing 10x to 100x increase in atom flux to lower sensor noise (atom shot noise).
ii) Development of more powerful lasers with faster pulses to increase the splitting of the atomic wavefunction for enhanced sensitivity.
iii) Operation and control of multiple atom sources to increase frequency coverage and sensitivity.
iv) Improve cold atom preparation cycle to support multiplexed interferometers (multiple simultaneous interferometers) to increase sampling rate and sensitivity.

In addition to benefiting atomic gravitational-wave detectors, these R\&D efforts have the potential to enhance the performance of atom and traditional laser interferometers in other quantum sensing applications, such as inertial sensing.

% One example is quantum optimal control R\&D being carried out at these centers which could lead to atom optics pulses with higher efficiency and better robustness to noise, helping our efforts to achieve larger momentum transfer atom optics. Significant R\&D will also be required to developed advanced quantum sensing methods such as spin squeezing and increasing large-momentum-transfer techniques. Other items include:
% \begin{itemize}
%     \item Improving cooling of atoms by developing 10x to 100x increase in atom flux to lower sensor noise (atom shot noise)
%     \item Development of more powerful lasers with faster pulses to increase the splitting of the atomic wavefunction for enhanced sensitivity
%     \item Operation and control of multiple atom sources to increase frequency coverage and sensitivity
%     \item Improve cold atom preparation cycle to support multiplexed interferometers (multiple simultaneous interferometers) to increase sampling rate and sensitivity
% \end{itemize}

\section{Future possibilities: Lunar-based Gravitational-Wave Detectors}
\label{sec:lunar}

One of the most challenging frequency range to measure gravitational waves is from deci-Hz to 1~Hz. This range is too low for the proposed Earth-based gravitational-wave detectors Cosmic Explorer \cite{Reitze:2019iox} and Einstein Telescope \cite{Punturo:2010zz} due to Newtonian noise limitations (section \ref{sec:NN}), and too high for the LISA space mission~\cite{Danzmann_1996}. Lunar seismic data from the Apollo era has in the past been used to to directly look for gravitational-waves in that band\cite{PhysRevD.90.102001,Harms_2021}, using the entire moon as a detector. Looking further into the future, 
%The universe offers a rich set of astrophysical sources in this regime \cite{Mandel_2018}, whose observations will open unique tests of general relativity and physics beyond the Standard Model \cite{sedda2019missing}. In addition, this frequency window offers the best chance for directly observing primordial gravitational waves - lower frequencies are foreground-source confusion-limited, and at higher frequencies the expected primordial spectrum lies below the sensitivity range of planned detectors.
the proposed moon-based Gravitational-Wave Lunar Observatory for Cosmology (GLOC) may provide a unique access this deci-Hz gravitational-wave regime \cite{GLOC_2021}. The projected sensitivity of GLOC and its comparison with terrestrial and space-based detectors is shown in discussed in  \cite{GLOC_2021}.  
The Moon offers a possible environment for constructing a large-scale interferometer as a gravitational-wave detector. The seismometers left from the Apollo missions suggest that at low-frequencies ($0.1 {\sim} 5$ Hz) the seismic background on the Moon is three orders of magnitude lower than on Earth \cite{Hanada2005}, providing a large reduction in the low-frequency Newtonian noise. 
The atmospheric pressure on the surface of the Moon during sunrise is comparable to the currently implemented 8 km ultra high vacuum ($10^{-10}$ torr) at each of the LIGO facilities \cite{PhysRevD.102.062003, Johnson1972}.  The presence of vacuum just above Moon's solid terrain provides a great benefit in extending the LIGO interferometer length at minimal cost.
With the advent of NASA Artemis, NASA Commercial Crew Program and ESA’s European Large Logistics Lander project there is a possibility of returning to the Moon this decade. One of the science priorities for NASA Artemis is in utilizing the unique environment of Moon to study the universe and a fundamental physics project on the lunar surface could be envisioned in the future.

\section{Outlook}
\label{sec:ol}

The next two decades will see significant growth in gravitational-wave astronomy, astrophysics and cosmology. The sensitivity of the Advanced LIGO detectors and its international partners VIRGO and KAGRA will continue to be improved for at least the next 10 years. Beyond that, on a mid-2030's timescale, the next-generation projects Cosmic Explorer and Einstein Telescope will take gravitational-wave observations back in cosmic time to the remnants of the first stars. In parallel atom interferometer prototypes are developing the technology for a new way to access the gravitational-wave spectrum at 1~Hz.

In addition, the complementary coverage of the entire gravitational-wave spectrum will improve with the LISA mission taking flight in the 2030s, with improvements in the NANOGrav pulsar-timing network, and with the implementation of the partially DOE funded CMB-S4 project's search for primordial gravitational-waves. For the first time multi-band gravitational-wave observations will be possible. Such observations can take to form of binaries sweeping across multiple bands, allowing merger predictions, spectral consistencies and black hole population studies covering a wide mass range.

Gravitational-wave detectors address a number of DOE science priorities, including the study of dense matter in neutron stars, measurements of the cosmological acceleration and dark energy, possible access to gravitational-wave signatures from the early universe, and a new way for direct dark matter observations. Just as importantly, DOE laboratories have the technical expertise needed to address the challenges of next-generation detectors, including among others large-scale ultra-high vacuum systems and the operation of new large facilities.

Finally, we discussed the prospect of a future lunar-based gravitational-wave observatory, which would provide a uniquely low seismic background in a band critical for the direct observation of primordial gravitational waves.

%Lunar observatory [Karan Jani]

%How does space-based interferometer or atom interferometery relate to these ground based facilities

%Broad observation band (Nanograv, Lisa, AI, GW etc)
%Observation capability in all bands, with localization

%The list of endorsers is still in preparation.
%If you wish to endorse this white paper, please email Stefan Ballmer, sballmer@syr.edu by March 29, 2022.
\newpage
\section{Endorsers}

\begin{tabular}{ |p{2in}|p{4in}|  }
 \hline
 \multicolumn{2}{|c|}{Endorsers A-J} \\
 \hline
Name & Affiliation \\
 \hline
Sambaran Banerjee & University of Bonn \\
Enrico Barausse & SISSA (Trieste  Italy) \\
Barry C Barish & Caltech and UC Riverside \\
Edo Berger & Harvard University \\
Beverly K. Berger & retired \\
Emanuele Berti & Johns Hopkins University \\
GariLynn Billingsley & California Institute of Technology \\
Ofek Birnholtz & Bar-Ilan University \\
Floor Broekgaarden & Center for Astrophysics | Harvard, Smithsonian \\
Senem Çabuk & Ankara University \\
Craig Cahillane & LIGO Hanford Observatory \\
Mesut Caliskan & Johns Hopkins University \\
Marco Cavaglia & Missouri University of Science and Technology \\
Poonam Chandra & Tata Institute of Fundamental Research  India \\
Sanha Cheong & Stanford / SLAC \\
Cecilia Chirenti & University of Maryland \\
Giacomo Ciani & University of Padova (Italy) \\
Alessandra Corsi & Texas Tech University \\
Saurya Das & University of Lethbridge \\
Nicholas DePorzio & Harvard University \\
Riccardo DeSalvo & Riclab LLC \\
Tejas Deshpande & Northwestern University \\
Daniela Doneva & University of Tuebingen \\
Bruce Edelman & University of Oregon \\
Martin Fejer & Stanford University \\
Giacomo Fragione & Northwestern University \\
Raymond Frey & University of Oregon \\
Paul Fulda & University of Florida \\
Gianluca Gemme & Istituto Nazionale di Fisica Nucleare (INFN)  Italy \\
Oliver Gerberding & Universität Hamburg  Germany \\
Mandeep S. S. Gill & Stanford University \\
Gabriela Gonzalez & Louisiana State University \\
Daryl Haggard & McGill University \\
Jan Harms & Gran Sasso  Science Institute \\
Carl-Johan Haster & Massachusetts Institute of Technology \\
Martin Hendry & University of Glasgow \\
Charles Horowitz & Indiana University \\
James Hough & University of Glasgow \\
Bala Iyer & ICTS-TIFR  Bangalore \\
Shang-Jie Jin & Northeastern University  China \\
 \hline
\end{tabular}

\begin{tabular}{ |p{2in}|p{4in}|  }
 \hline
 \multicolumn{2}{|c|}{Endorsers K-R} \\
 \hline
Name & Affiliation \\
 \hline
Vassiliki Kalogera & Northwestern University \\
Keita Kawabe & LIGO Hanford Observatory  Caltech \\
Joey Shapiro Key & U Washington Bothell \\
Kostas Kokkotas & University of Tuebingen \\
Tim Kovachy & Northwestern \\
Philippe Landry & Canadian Institute for Theoretical Astrophysics \\
Paul Lasky & Monash University \\
Albert Lazzarini & Caltech \\
Sungho Lee & Korea Astronomy and Space Science Institute \\
Kyung Ha Lee & Sungkyunkwan University \\
luis lehner & perimeter institute \\
Ilya Mandel & Monash University \\
Vuk Mandic & University of Minnesota Twin Cities \\
Szabolcs Marka & Columbia University ing the City of New York \\
Rodica Martin & Montclair State University \\
Mario Martinez & ICREA/IFAE-Barcelona \\
Jess McIver & University of British Columbia \\
Cole Miller & University of Maryland \\
Edoardo Milotti & INFN - Italy \\
Jeremiah Mitchell & University of Cambridge \\
Guenakh Mitselmakher & University of Florida \\
Shinji MIYOKI & ICRR  The University of Tokyo \\
Sergei Nagaitsev & Fermilab/Uchicago \\
Rohit Nair & Model college \\
David A. Nichols & University of Virginia \\
Alexander H. Nitz & Max Planck Institute for Gravitational Physics  AEI \\
Rafael C. Nunes & Instituto Nacional de Pesquisas Espaciais - Brazil \\
Benjamin J Owen & Texas Tech University \\
Hiranya Peiris & University College London / Stockholm University \\
Harald P. Pfeiffer & Max Planck Institute for Gravitational Physics  AEI \\
Jorge Piekarewicz & Florida State University \\
Geraint Pratten & University of Birmingham \\
David Radice & Penn State \\
Jocelyn Read & California State University Fullerton \\
David Reitze & Caltech \\
Michael Rezac & Cal State Fullerton \\
Jonathan Richardson & University of California Riverside \\
Keith Riles & University of Michigan \\
Jan Rudolph & Stanford \\
 \hline
\end{tabular}

\begin{tabular}{ |p{2in}|p{4in}|  }
 \hline
 \multicolumn{2}{|c|}{Endorsers S-Z} \\
 \hline
Name & Affiliation \\
 \hline
Mairi Sakellariadou & King's College London \\
Gary H. Sanders & Simons Observatory  Simons Foundation \\
Misao Sasaki  & Kavli IPMU  University of Tokyo \\
Robert Schofield & University of Oregon \\
Lijing Shao & Kavli Institute for Astronomy and Astrophysics  Peking U. \\
Peter Shawhan & University of Maryland \\
Deirdre Shoemaker & University of Texas at Austin \\
David Shoemaker & MIT \\
Deirdre Shoemaker & University of Texas at Austin \\
Bram Slagmolen & Australian National University \\
Joshua Smith & California State University  Fullerton \\
Frank Soboczenski & King's College London \\
Yevgeny Stadnik & University of Sydney \\
Leo C. Stein & University of Mississippi \\
Jan Steinhoff & Albert Einstein Institute (Potsdam) \\
Nikolaos Stergioulas & Aristotle University of Thessaloniki \\
Dejan Stojkovic & University at Buffalo \\
Ling Sun & The Australian National University \\
Chen Sun & Tel-Aviv University \\
Nicola Tamanini & Laboratoire des 2 Infinis - Toulouse (L2IT/CNRS) \\
Satoshi Tanioka & Syracuse University \\
David Tanner & University of Florida \\
Elias C. Vagenas & Kuwait University \\
Gabriele Vajente & California Institute of technology \\
Flavio Vetrano & Urbino University - Italy \\
Alan J. Weinstein & Caltech \\
Rainer Weiss & MIT \\
Karl Wette & Australian National University \\
Hiroaki Yamamoto & Caltech \\
Stoytcho Yazadjiev & Sofia University \\
Chunnong Zhao & The University of Western Australia \\
 \hline
\end{tabular}

% References
\bibliographystyle{JHEP.bst}
\bibliography{main.bib}

\end{document}